\documentstyle[12pt]{article}

\begin{document}
\begin{titlepage}
\setcounter{page}{1}
\title{Vacuum polarization around a straight wire carrying a steady current}
\author{B. Linet \thanks{E-mail: linet@ccr.jussieu.fr}  \\
\mbox{\small Laboratoire de Gravitation et Cosmologie Relativistes} \\
\mbox{\small CNRS/URA 769, Universit\'{e} Pierre et Marie Curie} \\
\mbox{\small Tour 22/12, Bo\^{\i}te Courrier 142} \\
\mbox{\small 4, Place Jussieu, 75252 PARIS CEDEX 05, France}}
\maketitle

\begin{abstract}

Motivated by the example of the superconducting cosmic string which can be
a physical representation
of a straight wire carrying a steady current, we derive in this case
the explicit
expressions of the induced vector potential, current density and magnetic
field due to the vacuum polarization
at the first order in the fine structure constant.

{\em Classification: 0370}

\end{abstract}

\end{titlepage}

\section{Introduction}

Cosmic strings could be produced at a phase transition in the early universe
and have survived to the present day \cite{kibble,vilenkin}. Moreover,
Witten \cite{witten} has shown that cosmic strings can behave like
superconductors
and carry very large electric current up to $10^{21}$ A. So, they can
generate a very large electromagnetic field.

When the gravitational field is neglected, a static and straight
superconducting cosmic string is described in the Minkowsksi spacetime,
outside the core of the string, by an infinite straight
wire carrying a steady current. It generates a stationary magnetic field.
Recently, Amsterdamski and O'Connor \cite{amsterdamski} and also Mankiewicz
and Zambowicz \cite{mankiewicz} have studied the vacuum polarization around
this superconducting cosmic string. These authors use the
results established in the case of a strong but slowly varying magnetic field.
However, Amsterdamski and O'Connor \cite{amsterdamski} have furthermore
fulfilled an alternative approximation at the first order in a power
series expansion in the strength of the inducing magnetic field.

The present paper is concerned with this last approach of the determination
of the vacuum polarization around the superconducting cosmic string.
Indeed, when the pair
creation is neglected, the induced current resulting from the vacuum
polarization in an external current has been determined in general by
Serber \cite{serber} at the first
order in the fine structure constant $\alpha$. For a point charge at rest the
induced
electrostatic potential has been given by Uehling \cite{uehling} but
probably due to the lack of
physical motivation, so far as we know, the induced magnetic field has not
been calculated for this straight current. Now the straight superconducting
cosmic string
is an interesting example therefore we will determine the induced vector
potential, current density and magnetic field in the case of an infinite
straight wire carrying a steady current.

The plan is as follows. In section 2, we recall the formula giving the
induced vector potential. We apply in section 3 this formula in the case
of a straight current. We also give the induced current density.
We calculate the induced magnetic field in section 4 and we give the
screening of the intensity of the current in the vicinity of the straight
wire. We add in section 5 some concluding remarks.

\section{Vacuum polarization in an external steady current}

In inertial coordinates ($t,x^{k}$), the Maxwell equations for a steady
current $j^{i}$ which is conserved reduce to
\begin{equation}
\label{2.1}
\triangle A^{i}=j^{i} \quad {\rm with} \quad \partial_{i}A^{i}=0
\end{equation}
where $A^{i}$ is the vector potential. The induced current $<j^{i}>$ due to
the vacuum polarization in an external current $j^{i}$ has been determined
by Serber \cite{serber} at the first order in $\alpha$. But we use actually
the equivalent integral expression given by Schwinger \cite{schwinger}
\begin{equation}
\label{2.2}
<j^{i}(x^{k})>=-\frac{\alpha}{6\pi^{2}}\int K(\mid x^{k}-x'^{k}\mid )
\triangle j^{i}(x'^{k})d^{3}x'
\end{equation}
where the function $K$ has the expression
\begin{equation}
\label{2.3}
K(r)=\frac{1}{r}\int_{1}^{\infty} \exp (-2mr\xi
(\frac{1}{\xi^{2}}+\frac{1}{2\xi^{4}})(\xi^{2}-1)^{1/2}d\xi
\end{equation}
$m$ being the mass of the electron ($\hbar =c=1$). The induced vector
potential $<A^{i}>$ is then determined from the Maxwell equations with
current (\ref{2.2}). We derive immediately the formula
\begin{equation}
\label{2.4}
<A^{i}(x^{k})>=-\frac{\alpha}{6\pi^{2}}\int K(\mid x^{k}-x'^{k}\mid )
j^{i}(x'^{k})d^{3}x'
\end{equation}

\section{Case of a straight wire carrying a steady current}

We consider that the superconducting cosmic string is parallel to the $x^{3}$
axis and carries a steady current of intensity $I$. The non-vanishing
component of the electric current density along the $x^{3}$ axis is
\begin{equation}
\label{3.1}
j^{3}(x^{k})=I\delta (x^{1})\delta (x^{2})
\end{equation}
We now introduce the cylindrical coordinates ($\rho ,\varphi ,z$). The vector
potential $A^{i}$ generated by current (\ref{3.1}) has only the component
$A^{z}$ which has the expression
\begin{equation}
\label{3.2}
A^{z}(\rho )=\frac{I}{2\pi} {\rm ln}\frac{\rho}{\rho_{0}}
\end{equation}
where $\rho_{0}$ is an arbitrary length. By deriving $A^{z}$ with respect to
$\rho$, we obtain the magnetic field whose only non-zero
component is $B_{\varphi}$ given by
\begin{equation}
\label{3.3}
B_{\varphi}(\rho )=\frac{I}{2\pi\rho}
\end{equation}

The induced vector potential $<A^{i}>$ is given by formula (\ref{2.4})
with current (\ref{3.1}) in which we perform the integration with respect to
$x'^{1}$ and $x'^{2}$ to obtain
\begin{equation}
\label{3.4}
<A^{z}(x^{k})>=-\frac{\alpha I}{6\pi^{2}}\int_{-\infty}^{+\infty}
K(\sqrt{\rho^{2}+(z-z')^{2}}dz'
\end{equation}
where $A^{z}$ depends only on $\rho$ as expected. By inserting expression
(\ref{2.3}) of the function $K$ in (\ref{3.4}), we have the double integral
\begin{eqnarray}
\label{3.5}
\nonumber <A^{z}(\rho )>&=&-\frac{\alpha I}{3\pi^{2}}\int_{1}^{\infty}
\int_{0}^{\infty}
\frac{\ exp (-2m\sqrt{\rho^{2}+u^{2}}\xi)}{\sqrt{\rho^{2}+u^{2}}} \\
& &\times (\frac{1}{\xi^{2}}+\frac{1}{2\xi^{4}})(\xi^{2}-1)^{1/2}d\xi du
\end{eqnarray}

Now, we turn to express $<A^{z}>$ in terms of elementary functions.
For this purpose, we perform the change of variables $u=\rho \sinh y$
and $\xi =\cosh x$ in integral (\ref{3.5}). We thus get
\begin{eqnarray}
\label{3.7}
\nonumber <A^{z}(\rho )>&=&-\frac{\alpha I}{3\pi^{2}}\int_{0}^{\infty}
\int_{0}^{\infty} \exp (-2m\rho \cosh x \cosh y) \\
& &\times (1+\frac{1}{2\cosh^{2}x})\tanh^{2}xdxdy
\end{eqnarray}
Now the modified Bessel function $K_{0}$ can be expressed as
\begin{equation}
\label{3.8}
K_{0}(z)=\int_{0}^{\infty}\exp (-z \cosh y)dy
\end{equation}
therefore we can reduce expression (\ref{3.7}) in the
form of a simple integral expression involving $K_{0}$
\begin{equation}
\label{3.9}
<A^{z}(\rho )>=-\frac{\alpha I}{3\pi^{2}}\int_{0}^{\infty}
K_{0}(2m\rho \cosh x)[1-\frac{1}{2\cosh^{2}x}-\frac{1}{2\cosh^{4}x}]dx
\end{equation}
However, except the first integral appearing in (\ref{3.9}), we have not
found the two other integrals in the tables. In appendix A we give
expressions  (\ref{a1}), (\ref{a2}) and (\ref{a3}) of the necessary
integrals. In consequence, we can get an explicit expression for the induced
vector potential $<A^{i}>$ in terms of modified Bessel functions $K_{n}$
\begin{eqnarray}
\label{3.10}
\nonumber <A^{z}(\rho )>=-\frac{\alpha I}{3\pi^{2}}
\{&& \frac{1}{2}[K_{0}(m\rho )]^{2}
-\frac{5}{6}m\rho K_{0}(m\rho )K_{1}(m\rho ) \\
\nonumber & &-\frac{2}{3}m^{2}\rho^{2}[K_{0}(m\rho )]^{2}
+\frac{7}{9}m^{2}\rho^{2}[K_{1}(m\rho )]^{2}  \\
\nonumber & &-\frac{2}{9}m^{4}\rho^{4}[K_{0}(m\rho )]^{2}
+\frac{2}{9}m^{4}\rho^{4}[K_{1}(m\rho )]^{2} \\
& &-\frac{2}{9}m^{3}\rho^{3}K_{0}(m\rho )K_{1}(m\rho ) \quad \}
\end{eqnarray}
We see immediately from (\ref{3.10}) that the range of the induced vector
potential is $1/2m$.

To be complete, we also give the induced current density $<j^{i}>$. By using
the identity
\[
z^{2}K_{0}''(z)+zK_{0}'(z)-z^{2}K_{0}(z)=0
\]
we can derive $<j^{z}>$ from Maxwell's equations (\ref{2.1}). We find
\begin{eqnarray}
\label{3.11}
\nonumber <j^{z}(\rho )>&=&-\frac{4\alpha Im^{2}}{3\pi^{2}}\int_{0}^{\infty}
K_{0}(2m\rho \cosh x) \\
&&\times [\cosh^{2}x-\frac{1}{2}-\frac{1}{2\cosh^{2}x}]dx
\end{eqnarray}
We have verified that integral expression (\ref{3.11}) coincides with the
result of Amsterdamski and O'Connor \cite{amsterdamski} after a change of
variable. With integrals (\ref{a1}), (\ref{c1}) and (\ref{a2}) given in
appendix A, we get the explicit expression of the induced current density
in terms of modified Bessel functions
\begin{eqnarray}
\label{3.12}
\nonumber <j^{z}(\rho )>=-\frac{\alpha Im^{2}}{3\pi^{2}}
\{ &&[K_{1}(m\rho )]^{2}
-2m\rho K_{0}(m\rho )K_{1}(m\rho ) \\
&&-2m^{2}\rho^{2}[K_{0}(m\rho )]^{2}+2m^{2}\rho^{2}[K_{1}(m\rho )]^{2}
\quad \}
\end{eqnarray}
The current density $<j^{z}>$ vanishes exponentially for large $\rho$ and
diverges as $\alpha I/3\pi^{2}\rho^{2}$ at $\rho =0$ since
$K_{1}(z)\sim 1/z$.
\section{Screening of the current by vacuum polarization}

The modification due to the vacuum polarization of the external magnetic
field (\ref{3.3})
occurs mainly near the straight wire. We can directly calculate the induced
magnetic field $<B_{\varphi}>$ by deriving expression (\ref{3.10})
with respect to $\rho$. Alternatively, we can also derive its integral
expression
\begin{eqnarray}
\label{4.1}
\nonumber <B_{\varphi}(\rho )>&=&\frac{2\alpha Im}{3\pi^{2}}\int_{0}^{\infty}
K_{1}(2m\rho \cosh x) \\
&&\times [\cosh x-\frac{1}{2\cosh x}-\frac{1}{2\cosh^{3}x}]dx
\end{eqnarray}
where we have use the fact that $K_{0}'=-K_{1}$. In appendix A we give
expressions (\ref{b1}), (\ref{b2}) and (\ref{b3}) of the necessary integrals.
Thus, we can get the explicit expression
of the induced magnetic field $<B^{i}>$ in terms of modified Bessel functions
\begin{eqnarray}
\label{4.2}
\nonumber <B_{\varphi}(\rho )>=\frac{\alpha Im}{3\pi^{2}}
\{&&K_{0}(m\rho )K_{1}(m\rho )
+\frac{1}{2}m\rho [K_{0}(m\rho )]^{2} \\
\nonumber & &-\frac{5}{6}m\rho [K_{1}(m\rho )]^{2}+
\frac{2}{3}m^{2}\rho^{2}K_{0}(m\rho )K_{1}(m\rho ) \\
& &+\frac{2}{3}m^{3}\rho^{3}[K_{0}(m\rho )]^{2}-
\frac{2}{3}m^{3}\rho^{3}[K_{1}(m\rho )]^{2} \quad \}
\end{eqnarray}
The total magnetic field $B^{i}$ generated by a straight current is
\begin{equation}
\label{4.2a}
B_{\varphi}(\rho )=\frac{I}{2\pi \rho}+<B_{\varphi}(\rho )>
\end{equation}
when the vacuum polarization is taken into account at the first order
in $\alpha$.

We are now in a position to work out the screening of the source of the
magnetic field.
This procedure is most significant that the analyse of the induced current
density given by (\ref{3.12}). From the asymptotic formulas
\[
K_{0}(z)\sim -\gamma -{\rm ln}(z/2) \quad {\rm and}
\quad K_{1}(z)\sim \frac{1}{z}
\]
where $\gamma$ is Euler's constant,
we obtain easily from (\ref{4.2}) the asymptotic form of $<B_{\varphi}>$ for
small values of $m\rho$
\begin{equation}
\label{4.3}
<B_{\varphi}(\rho )>\sim \frac{\alpha I}{3\pi^{2}\rho}(-\gamma -\frac{5}{6}
+{\rm ln}2-{\rm ln}m\rho )
\end{equation}
With formula (\ref{4.2a}), we evaluate easily the screening of the intensity
the current in the vicinity of the straight wire
\begin{equation}
\label{4.4}
I_{eff}\approx I[1+\frac{\alpha}{3\pi^{2}}(-\gamma -\frac{5}{6}+{\rm ln}2-
{\rm ln}m\rho )]
\end{equation}
for small values of $m\rho$.

\section{Conclusion}

Our main result is the explicit determination of induced quantities: vector
potential (\ref{3.10}), current density (\ref{3.12}) and magnetic field
(\ref{4.2}), due to the vacuum polarization
in an external straight current. However, these results are valid in domains
where the magnetic field is not too strong
because we have not considered higher order effects in the fine
structure constant. In the vicinity of this straight wire we have found
asymptotic expansion (\ref{4.3}) of the induced magnetic field.
For above mentioned reasons, we cannot consider this formula for distances
from string too small. Furthermore, vacuum polarization effects resulting
from other charged particles with mass greater than the electron mass
can eventually occur.
\appendix
\section{Appendix}

We write down a general formula \cite{table} that we will be needed in the
determination of our integrals
\[
\int_{0}^{\infty}K_{2\mu}(2z\cosh x)
\cosh 2\nu xdx=\frac{1}{2}K_{\mu+\nu}(z)K_{\mu-\nu}(z) \quad {\rm for}
\quad {\cal R}e(z)>0
\]
{}From this we have immediately the following results
\begin{equation}
\label{a1}
\int_{0}^{\infty}K_{0}(b\cosh x)dx=\frac{1}{2}[K_{0}(b/2)]^{2}
\end{equation}
\begin{equation}
\label{b1}
\int_{0}^{\infty}K_{1}(b\cosh x)\cosh xdx=\frac{1}{2}K_{0}(b/2)K_{1}(b/2)
\end{equation}
and after the use of the relation
\[
\cosh^{2}x=\frac{1}{2}(1+\cosh 2x)
\]
\begin{equation}
\label{c1}
\int_{0}^{\infty}K_{0}(b\cosh x)\cosh^{2}xdx=\frac{1}{4}[K_{0}(b/2)]^{2}
+\frac{1}{4}[K_{1}(b/2)]^{2}
\end{equation}

The second integral that we have needed can be performed as follows.
After an integration by part, it becomes
\[
\int_{0}^{\infty}\frac{K_{0}(b\cosh x)}{b^{2}\cosh^{2}x}dx=
\frac{1}{b}\int_{0}^{\infty}dx
K_{1}(b\cosh x)\cosh x-\int_{0}^{\infty}dx\frac{K_{1}(b\cosh x)}{b\cosh x}
\]
by using the relation $K_{0}'=-K_{1}$.
By virtue of the recurrence relation
\[
zK_{\nu -1}(z)-zK_{\nu +1}(z)=-2\nu K_{\nu}(z)
\]
we get moreover
\begin{equation}
\label{b2}
\int_{0}^{\infty}\frac{K_{1}(b\cosh x)}{\cosh x}dx=-
\frac{b}{4}[K_{0}(b/2)]^{2}+
\frac{b}{4}[K_{1}(b/2)]^{2}
\end{equation}
Finally by combining these results, we obtain thereby the expression of the
second integral
\begin{equation}
\label{a2}
\int_{0}^{\infty}\frac{K_{0}(b\cosh x)}{\cosh^{2}x}dx=
\frac{b}{2}K_{0}(b/2)K_{1}(b/2)
+\frac{b^{2}}{4}[K_{0}(b/2)]^{2}-\frac{b^{2}}{4}[K_{1}(b/2)]^{2}
\end{equation}

The method of computing the third integral is the same. So,
we will not reproduce the calculations.
We will have to know the expression of the following integral
\begin{eqnarray}
\nonumber \int_{0}^{\infty}\frac{K_{1}(b\cosh x)}{\cosh^{3}x}dx&=&-
\frac{b^{2}}{4}K_{0}(b/2)K_{1}(b/2)
-\frac{3b^{3}}{32}[K_{0}(b/2)]^{2} \\
\nonumber & &+\frac{b^{3}}{12}[K_{1}(b/2)]^{2}+
\frac{b^{3}}{96}[K_{2}(b/2)]^{2}
\end{eqnarray}
that we can transform in the form
\begin{eqnarray}
\label{b3}
\nonumber \int_{0}^{\infty}\frac{K_{1}(b\cosh x)}{\cosh^{3}x}dx&=&
-\frac{b^{2}}{6}K_{0}(b/2)K_{1}(b/2)-\frac{b^{3}}{12}[K_{0}(b/2)]^{2} \\
& &+\frac{b^{3}}{12}[K_{1}(b/2)]^{2}+\frac{b}{6}[K_{1}(b/2)]^{2}
\end{eqnarray}
With this result and after some manipulations we will obtain the
expression of the third integral
\begin{eqnarray}
\label{a3}
\nonumber \int_{0}^{\infty}\frac{K_{0}(b\cosh x)}{\cosh^{4}x}dx&=&
\frac{b^{4}}{36}[K_{0}(b/2)]^{2}
-\frac{b^{4}}{36}[K_{1}(b/2)]^{2} \\
& &+\frac{b^{2}}{12}[K_{0}(b/2)]^{2}-\frac{5b^{2}}{36}[K_{1}(b/2)]^{2}  \\
\nonumber & &+\frac{b^{3}}{18}K_{0}(b/2)K_{1}(b/2)+
\frac{b}{3}K_{0}(b/2)K_{1}(b/2)
\end{eqnarray}


\newpage

\begin{thebibliography}{99}
\bibitem{kibble} Kibble T W B 1976 {\em J. Phys. A: Math. Gen.} {\bf 9} 1387
\bibitem{vilenkin} Vilenkin A 1985 {\em Phys. Rep.} {\bf 121} 263
\bibitem{witten} Witten E 1985 {\em Nucl. Phys. B} {\bf 249} 557
\bibitem{amsterdamski} Amsterdamski P and O'Connor D 1988 {\em Nucl. Phys. B}
{\bf 298} 429
\bibitem{mankiewicz} Mankiewicz L and Zembowicz R 1988 {Phys. Lett. B}
{\bf 202} 493
\bibitem{serber} Serber R 1935 {\em Phys. Rev.} {\bf 48} 49
\bibitem{uehling} Uehling E A 1935 {\em Phys. Rev.} {\bf 48} 55
\bibitem{schwinger} Schwinger J 1949 {\em Phys. Rev.} {\bf 75} 651
\bibitem{table} Prudnikov A P, Brychkov Yu A and Marichev O I 1986
{\em Integrals and Series vol. 2} (Gordon and Breach Science Publishers)
p. 358

\end{thebibliography}
\end{document}